\documentclass[5p, twocolumn, 9pt]{elsarticle}

\usepackage{multicol}
\usepackage{caption }

\usepackage{upgreek}
\usepackage{graphicx}
\usepackage{bm}
\usepackage{amsmath, amsthm, amssymb}
\usepackage{latexsym}
\usepackage{dcolumn}

\usepackage{mathtools}

\DeclarePairedDelimiter\ket{\lvert}{\rangle}
\DeclarePairedDelimiterX\braket[2]{\langle}{\rangle}{#1 \delimsize\vert #2}

\journal{JMR}

\begin{document}

\begin{frontmatter}



\title{Dynamic nuclear polarisation of liquids at one microtesla using circularly polarised RF with application to millimetre resolution MRI}

\author[1]{Ingo Hilschenz}
\author[1]{Sangwon Oh}
\author[1]{Seong-Joo Lee}
\author[1]{Kwon Kyu Yu}
\author[1,2]{Seong-min Hwang}
\author[1,2]{Kiwoong Kim}
\author[1,2,*]{Jeong Hyun Shim}

\address[1]{Korean Research Institute of Standards and Science, Daejeon 34113, Republic of Korea}
\address[2]{Department of Medical Physics, University of Science and Technology, Daejeon 34113, Republic of Korea}
\address[*]{jhshim@kriss.re.kr}

\begin{abstract}
Magnetic resonance imaging in ultra-low fields is often limited by mediocre signal-to-noise ratio hindering a higher resolution. Overhauser dynamic nuclear polarisation (O-DNP) using nitroxide radicals has been an efficient solution for enhancing the thermal nuclear polarisation. However, the concurrence of positive and negative polarisation enhancements arises in ultra-low fields resulting in a significantly reduced net enhancement, making O-DNP far less attractive. Here, we address this issue by applying circularly polarised RF. O-DNP with circularly polarised RF renders a considerably improved enhancement factor of around 150,000 at 1.2~$\upmu$T. A birdcage coil was adopted into a ultra-low field MRI system to generate the circularly polarised RF field homogeneously over a large volume. We acquired an MR image of a nitroxide radical solution with an average in-plane resolution of 1~mm. De-noising through compressive sensing further improved the image quality.
\end{abstract}


\end{frontmatter}


\section{Introduction}
\label{Introduction}

Nuclear magnetic resonance (NMR) at ultra-low fields has attained interest in several fields such as combining magnetic resonance imaging (MRI) with magnetoencephaolography (MEG) \cite{volegov2004simultaneous, vesanen2013hybrid}, resolving $J$-coupling structures in molecules \cite{McDermott2002, Ledbetter2009, Theis2013}, enhancing MRI contrast between certain biological tissues \cite{Sarah2012}, obtaining MR image of samples enclosed in metal \cite{Moessle2006}, detecting neuronal currents directly \cite{Kim2014,korber2016squids} and measuring fundamental quantities \cite{Griffith2009}. The detection of NMR signals in ultra-low fields, typically below the earth's magnetic field, requires highly sensitive sensors, such as superconducting quantum interference devices (SQUID) \cite{greenberg1998application, McDermott2004} or optically-pumped atomic magnetometers \cite{kominis2003, Budker2007}. However, the thermal nuclear polarisation in ultra-low fields is not sufficiently large. Thus, most implementations of ultra-low field NMR/MRI temporally increase the magnetic field above tens of millitesla through a process known as pre-polarisation. This has been conducted using either \textit{in situ} electromagnets or \textit{ex situ} permanent magnets. Strong pre-polarisation fields produced by electromagnets entail inducing eddy currents in the shielding walls and magnetise the superconducting pick-up coils, which may interfere with the detection of weak NMR signals \cite{Hwang2011,nieminen2011avoiding,zevenhoven2014,storm2016modular}. \textit{Ex situ} pre-polarisation requires the pneumatic shuttling of a sample from a remote location to the detection space. The shuttling can be an obstacle for reliable repetitions required in two-dimensional spectroscopy or MRI.

\begin{figure*}[t!]
\centering
\includegraphics[origin=c,clip=true,width=0.95\textwidth]{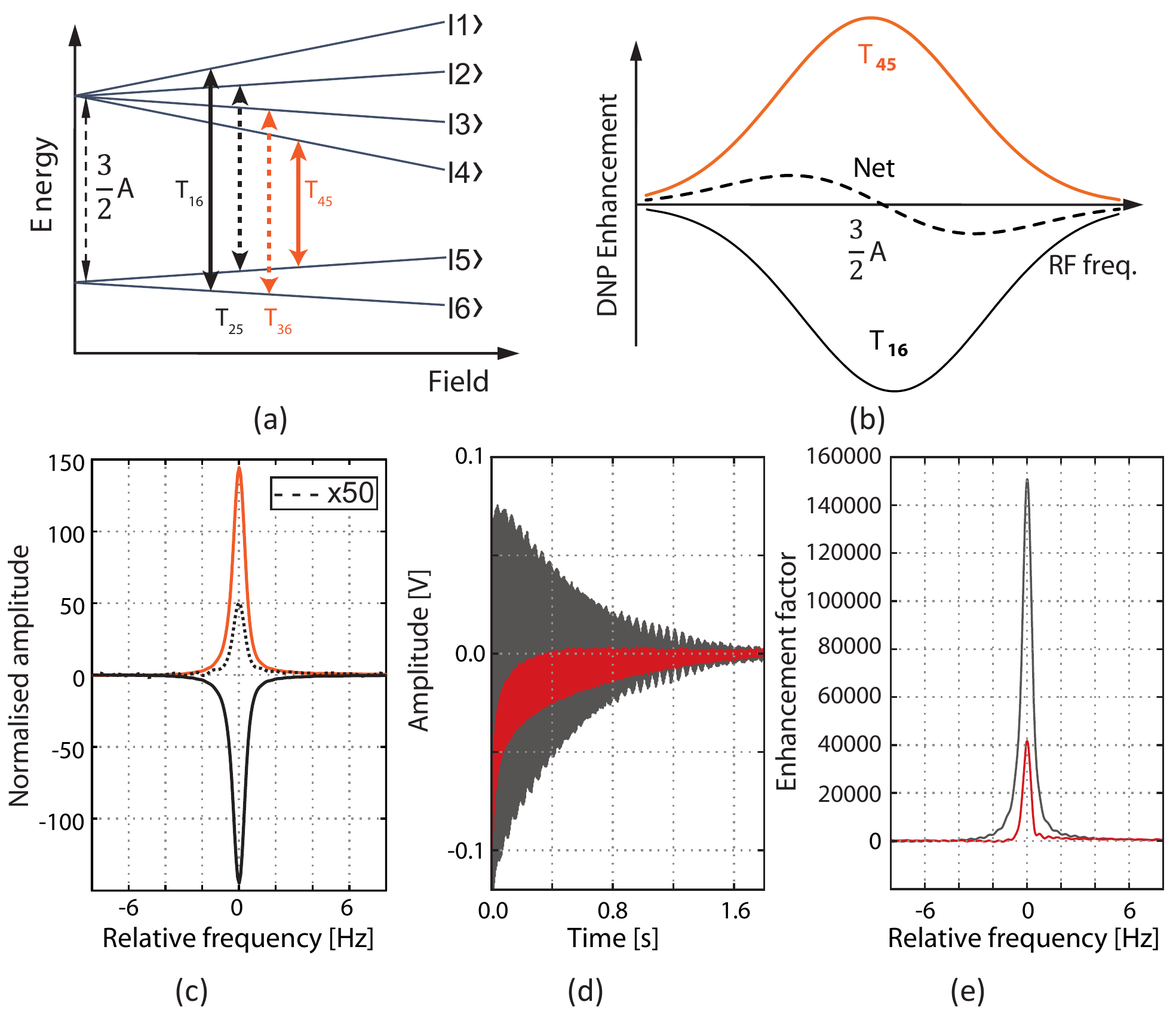}
\caption{(a) 	Energy levels of nitroxide radicals in ultra-low fields plotted as a function of the magnetic field. The allowed
							transitions when the magnetic field and oscillating RF field are orthogonal, are illustrated as arrows.
				 (b) 	A simulation of the Overhauser-enhanced nuclear polarisations for T$_{45}$ and T$_{16}$. Owing to the concurrence of 	
							positive (orange) and negative (black) enhancements, the net (dashed) is significantly reduced. We exaggerated it
							for the plot 50-fold for visibility, it symbolises as well what signal enhancement is to be expected when utilising a
							linear polarised RF. T$_{25}$ and T$_{36}$ have negligible contributions to the net enhancement (see the text).
			   (c) 	NMR spectra with a polarising field of $B_p$~1.2~$\upmu$T enhanced using O-DNP with circular polarisation
					    $\phi~=~90~^{\circ}$ (orange), $\phi = 270~^{\circ}$ (black). The vertical scale is normalised to the linearly
							polarised signal (dashed line). The RF frequency of the linear polarisation was 69~MHz which is near the maximum of
							the net curve in (b)\cite{lee2015dynamic} corresponding to what can be achieved with applying a circluar polarising
							field.
				 (d) 	NMR signals obtained with O-DNP using circularly polarised RF~(gray) and a pre-polarisation field of 50~mT~(red), and
				 (e) 	their spectra scaled to the enhancement in comparison with the thermal nuclear polarisation at 1.2~$\upmu$T.}
\label{fig:1}
\end{figure*}

 Inducing hyperpolarised states in the target samples has been an alternative approach~\cite{lee2015dynamic,Zotev2010,lurie1988proton,  Hoevener2013, Buckenmaier2017}. In Overhauser dynamic nuclear polarisation (O-DNP), the polarisation of electron spins belonging to free radicals is transferred to that of nuclear spins of the target molecules in a solution\cite{slichter2013principles}. However, O-DNP has hardly been used at a magnetic field range below a few millitesla \cite{ Zotev2010,krishna2002overhauser, sarracanie2014high, waddington2018overhauser}, because its effectiveness at enhancing the thermal polarisation significantly degrades in ultra-low fields. When applying the RF for O-DNP, positive and negative enhancements occur simultaneously, and in turn cancel each other out \cite{lee2015dynamic, Guiberteau1996}.
\\\indent Nevertheless, Lee \textit{et al.} \cite{lee2015magnetic} adopted O-DNP for MRI in a magnetic field of 34.5~$\upmu$T, unfortunately the image resolution was not satisfactory owing to the low enhancement of the thermal polarisation. In this study, we address this issue by using circularly polarised RF fields, which selectively induces either positive or negative enhancements. The thermal nuclear polarisation was increased by a factor of 150,000 at 1.2 $\upmu$T, overcoming the Overhauser limit. An Overhauser-enhanced MRI of a nitroxide radical solution with circularly polarised RF rendered a decent signal-to-noise ratio and allowed us to strengthen the field gradients for improving the image resolution. A birdcage coil, widely used in high-field MRI, serves as a transmitting coil for generating a circularly polarised RF with a satisfactory spatial uniformity. In addition, a compressed sensing technique was applied to de-noise the obtained images. This work demonstrates that higher image resolution can be obtained using O-DNP with circular polarised RF fields in combination with nitroxide radicals at the microtesla scale, without the aid of a strong pre-polarisation fields or shuttling.

\section{Theoretical model}
\label{Background theory}
To explain O-DNP in ultralow fields, we formulate a model and thereby express the enhancement factor in a more intuitive form, by neglecting the effects of a magnetic field when necessary.

\subsection{Spin Hamiltonian}
In the presence of a magnetic field $\bm{B}$, the Hamiltonian of a nitroxide radical can be expressed as
${\mathcal{H}}=A\bm{S}\cdot \bm{I} - (\gamma_e \bm{S}+\gamma_n \bm{I})\cdot\bm{B}$, in which $\bm{S}$ and $\bm{I}$ are electron ($S=1/2$) and $^{14}$N nuclear ($I=1$) spins. The gyromagnetic ratios of the electron and nuclear spins are labelled as $\gamma_e$ and $\gamma_n$, respectively. In ultra-low fields, the hyperfine coupling $A$ between electron and $^{14}$N spins in the radical is dominant over the Zeeman energies, making $\bm{S}$ and $\bm{I}$ strongly coupled. To diagonalise the Hamiltonian, the basis states $\ket{F,m_F}$ of spin $\bm{F}=~\bm{S}+\bm{I}$ are required. The Hamiltonian transforms into

\begin{equation}
		\mathcal{H}=A\frac{F(F+1)}{2}  - \gamma_F(F)\bm{F}\cdot \bm{B}+ C.
\end{equation}

The first term represents the zero field splitting (ZFS) of $3A/2$ between the quartet of $F=3/2$ and the doublet of $F=1/2$ in the energy diagram (Fig.~\ref{fig:1}~(a)). The second term is the Zeeman energy of the spin $\bm{F}$, and the last a constant independent of $F$. The exact expression of the gyromagnetic ratio $\gamma_F (F)$ is given as,

\begin{eqnarray}
\gamma_F (F) = &&\frac{\gamma_e}{2} \left[1+ \frac{S(S+1)-I(I+1)}{F(F+1)}\right] \nonumber \\
&+& \frac{\gamma_n}{2}\left[1 -  \frac{S(S+1)-I(I+1)}{F(F+1)}\right].\label{Eq_gamma_F}
\end{eqnarray}

We can approximate $\gamma_F(3/2) = \gamma_e/3 $ and $\gamma_F (1/2) = -\gamma_e/3$ as $\left|\gamma_e\right|$ is nearly 10$^4$ times larger than $\gamma_n$ ($\frac{\gamma_e}{2\pi}$=-28 GHz/T, $\frac{\gamma_n}{2\pi}$ = 3.077 MHz/T for $^{14}$N). The eigenstates depicted in Fig.~\ref{fig:1}(a) are $\ket{1}= \ket{3/2,3/2}$, $\ket{2}=\ket{3/2,1/2}$, $\ket{3}=\ket{3/2,-1/2}$, $\ket{4}=\ket{3/2,-3/2}$, $\ket{5}=\ket{1/2,-1/2}$ and $\ket{6}=\ket{1/2,1/2}$.

\subsection{Enhancement factor}
When an oscillating field $\bm{B}(t)$ is applied, the transitions between the states can be induced. The polarising field $B_p$ is perpendicular to $\bm{B}(t)=B_{RF}\left[\cos\left( \omega t\right)\hat{x}+\cos\left( \omega t+\phi\right)\hat{z}\right]$. Thus, T$_{16}$, T$_{25}$, T$_{36}$ and T$_{45}$, shown as arrows in Fig.~\ref{fig:1}~(a), are the allowed transitions under the selection rule $\Delta m_F=\pm 1$ and have non-negligible frequencies across ZFS. At one microtesla, the difference between the transition frequencies and ZFS is in the order of $\left|\gamma_F B \right|$$\simeq$ 10 kHz, which is insignificant compared to ZFS. Therefore, the four transitions occur simultaneously when exposed to an RF frequency of ZFS (72~MHz). Given the saturation of a transition T$_{ij}$, the DNP enhancement factor is proportional to the variation of the spin moment $\Delta\langle\gamma_F(F)\bm{F}_z \rangle_{ij}$. Thus, the formula for the enhancement factor $\varepsilon$ in Ref.~\cite{Guiberteau1996} can be applied to the spin $\bm{F}$ as,

\begin{equation}
\varepsilon_{ij} =  \rho f s  \frac{\Delta\langle\gamma_F(F)\bm{F}_z \rangle_{ij}} {|\gamma_F (F)|I_0},\label{Eq_Enhancement}
\end{equation}
in which $I_0$ is the proton spin polarisation in thermal equilibrium. ($\rho$, $f$, and $s$ are coupling constant, leakage factor, and saturation factor.)

According to the calculation (see the \ref{app:b}), the max enhancement for a transition T$_{ij}$ can be obtained as
\begin{equation}
\varepsilon_{ij}^{max} = -\frac{A}{4 \gamma_I B} \frac{m_F^i + m_F^j}{2},\label{Max_enhancement}
\end{equation}
where $m_F^i$ represents $m_F$ of the state $\ket{i}$. Thus, T$_{25}$ and T$_{36}$ do not exhibit significant enhancements, whereas T$_{16}$ and T$_{45}$ do, although with opposing signs. This leads to the concurrence of positive and negative enhancements, cancelling each other out. The net enhancement is significantly lower, when applying a linear polarised RF field, than the individual enhancements, as shown in Fig.~\ref{fig:1}~(b). Applying circularly polarised RF fields can help address this issue by selectively inducing either T$_{16}$ $(\Delta m_F =+1)$ or T$_{45}$ $(\Delta m_F = -1)$.

\begin{figure*}[t!]
\centering
\includegraphics[origin=c,clip=true,width=1\textwidth]{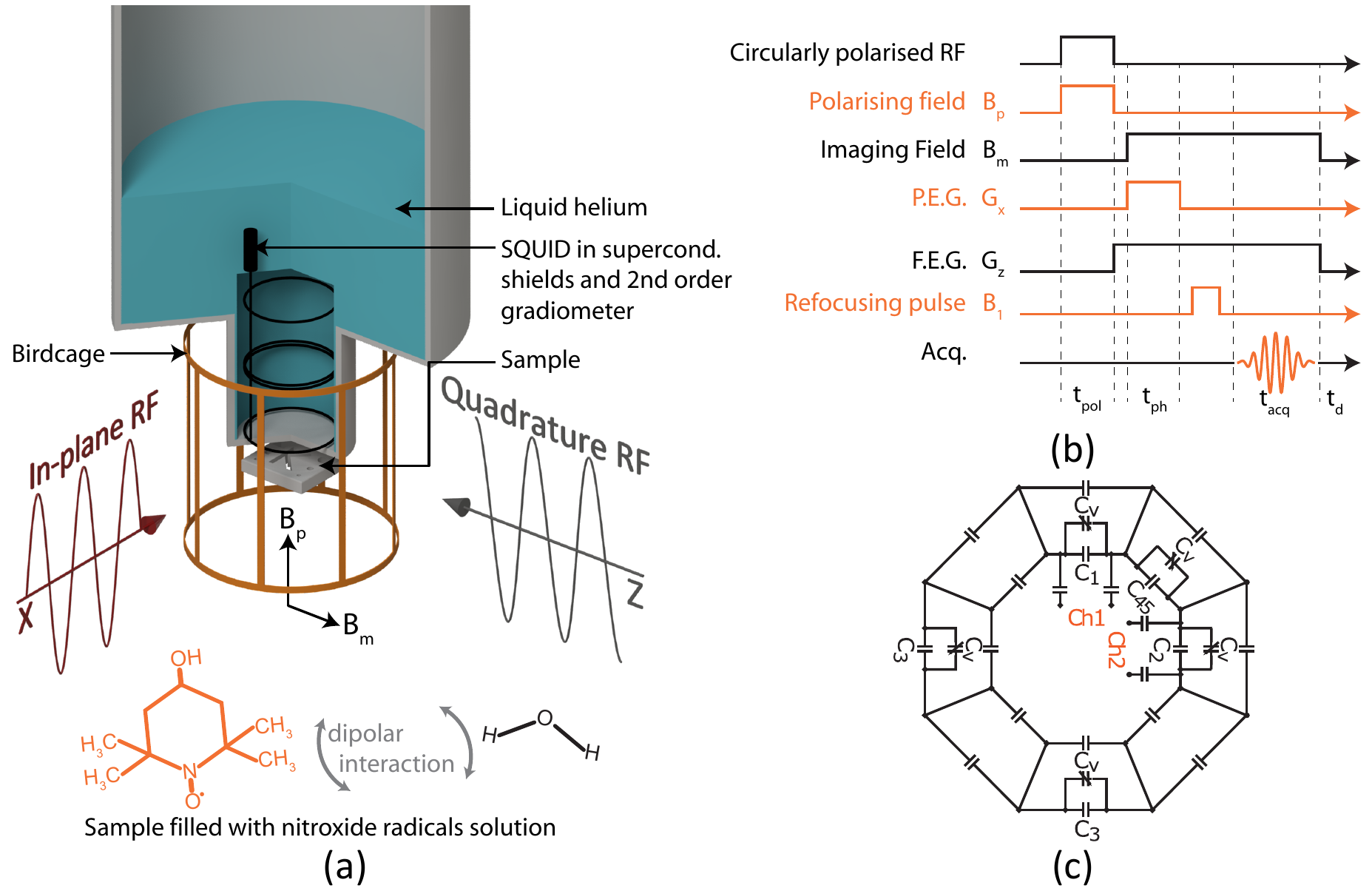}
\caption{	(a) Schematic of the ultra-low field MRI system, including the SQUID sensor, birdcage RF coil, and phantom that contains
							a nitroxide radical (TEMPOL) solution.
          (b) Timing sequence generating the spin-echo signals, with a refocusing $B_1$ pulse. Prior to this, the polarising field
					    $B_{pol}$ and circularly polarised RF are applied simultaneously for O-DNP. Phase-encoding (PEG) and frequency-
							encoding gradient fields (FEG) are coding the 2D image. We used a polarisation time $t_{pol}$ of 2000~ms, a phase
							encoding time $t_{ph}$ of 100~ms, and a read-out time $t_{acq}$ of 200~ms and the delay $t_{d}$ between consecutive 
							measurements is 1000~ms .
			    (c) Circuit schematic of the high-pass birdcage coil, the capacitor values of which are all approximately 50~pF, and the
							trimmer capacitors range is $C_V$~=~(1-12)~pF.}
\label{fig:2}
\end{figure*}

\section{Results}
\label{Results}
\subsection{O-DNP enhancement with circularly polarised RF}
Figure \ref{fig:1}~(c) discloses that, O-DNP with circularly polarised RF renders a considerably increased enhancement, over 140-fold. The dashed line corresponds to a data set acquired with linearly polarised RF, whereas the orange and black lines represent circularly polarised RF. The polarity of the O-DNP enhancement is solely determined based on the relative phase $\phi$ of the quadrature RF field (the orange ($\phi~=~90~^{\circ}$) and black ($\phi = 270~^{\circ}$) lines in Fig.~\ref{fig:1}~(c)), which is what the theoretical model predicts.
\\\indent In order to estimate how much O-DNP can enhance the thermal nuclear polarisation with a circularly polarised RF, at a polarising field of 1.2$~\upmu$T, we instead applied a pre-polarisation field $B_p$ of 50~mT for comparison. Figure~\ref{fig:1}~(d) illustrates that the higher polarising field increases the eddy currents, resulting in a notable dc-drift in the initial time after the pre-polarisation field has turned off (please compare the red and black curve). The Overhauser-enhanced NMR signal was found out to be 3.6~times as high as that obtained with the pre-polarisation field (Fig.~\ref{fig:1}~(d) and (e)). Thus, utilising O-DNP with circularly polarised RF, is equivalent to applying a pre-polarisation field of 180~mT, indicating the total enhancement factor can be estimated to around 150,000. The enhancement factor of the linearly polarised RF shown in Fig.~\ref{fig:1}~(b) is around 1030 at 1.2~$\upmu$T.

\begin{figure*}[t]
\centering
\includegraphics[origin=c,clip=true,width=1\textwidth]{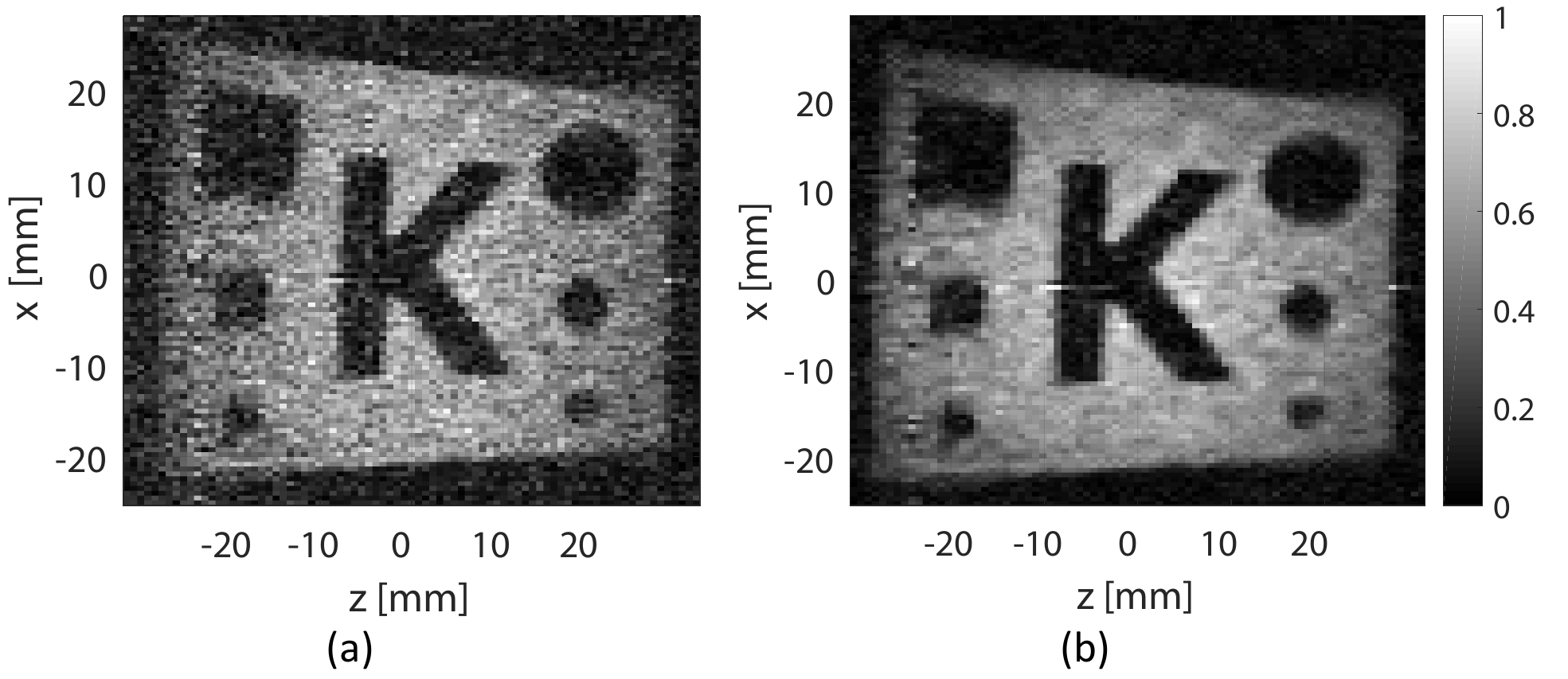}
\caption{Images of our phantom obtained at a voxel size of 0.76~mm~$*$~0.54~mm in the z- and x-directions:
		(a) unprocessed image, and
		(b) image from the same data set but treated using a compressed sensing method.}
\label{fig:3}
\end{figure*}

The degree of the enhancement with circularly polarised RF is far higher than what the conventional theory of O-DNP states when the Zeeman energy is dominant over the hyperfine coupling. According to it, the enhancement factor is limited to the ratio of gyromagnetic constants as $\frac{\gamma_S}{\gamma_I}$, which leads to only 660 for proton spins\cite{slichter2013principles}. According to Eq.~(\ref{Max_enhancement}), the enhancement factor can reach up to 235,000 at 1.2~$\upmu$T, which is certainly consistent with our result. Zero Field Splitting plays a crucial role for such a high enhancement factor. Due to its presence, the population difference between the upper quartet and lower doublet in Fig.~\ref{fig:1}~(a) becomes independent of the magnetic field and proportional to the ZFS as $\frac{3A}{2k_B T}$ (Eq.~(\ref{EqB3})), making the enhancement of the thermal nuclear polarisation scales as $\frac{A}{4 \gamma_I B}$ (Eq.~(\ref{Max_enhancement})). This elucidates how the enhancement beyond the Overhauser limit was accomplished and can be even higher at a lower magnetic field.

\subsection{Overhauser-enhanced MRI}
Overhauser-enhanced imaging was conducted using our ultra-low field MRI system (Fig.~\ref{fig:2}~(a)). The MR sequence is based on a spin-echo measurement (Fig. \ref{fig:2}~(b)) where we used a polarisation time $t_{pol}$ of 2000~ms, a phase encoding time $t_{ph}$ of 100~ms, a read-out time $t_{acq}$ of 200~ms and a delay time $t_d$ between measurements of 1000~ms. The frequency encoding gradient (FEG) $G_z$ strength was 229.8~~$\frac{\upmu \mathrm{T}}{\mathrm{m}}$, and the phase encoding gradient (PEG) $G_x$ was 186.0~$\frac{\upmu \mathrm{T}}{\mathrm{m}}$ at a maximum of 141~steps in-~between. The total acquisition time constituted to 77~min. The phantom in the present work is the same as in our previous study \cite{lee2015magnetic} and was filled with a 2~mM TEMPOL (4-hydroxy-2,2,6,6-tetramethylpiperidin-1-oxyl) solution which seemed to achieve the optimal result in the microtesla field range\cite{liebgott2003proton,nicholson1994application}. It has the dimensions of 54~mm~$*$~43.5~mm~$*$~40~mm, featuring a letter K in the middle as well as rectangles and circles with varying sizes (top 10~mm, middle 5~mm, bottom 3~mm) on the left and right sides, respectively. For O-DNP, we applied a polarising field $B_p$ of 1.2~$\upmu$T. The imaging field $B_{m}$ is 20.69~$\upmu$T, corresponding to a Larmor frequency of 880~Hz.
\\\indent The obtained image with a sub-millimetre voxel size (in the 2D plane) of 0.76~mm~$*$~0.56~mm, is shown in Fig.~\ref{fig:3}~(a). The image shows distortion owing to the presence of concomitant gradients \cite{nieminen2010solving} and inhomogeneity in the signal strength. The latter is caused by the spatial inhomogeneity in the sensitivity of the pick-up coil, as discussed in Ref.~\cite{lee2015dynamic}, and the off-resonance effect of the refocusing $B_1$ pulse (see the discussion).

\subsection{De-noising using compressive sensing}
To further improve the image quality, we applied compressive sensing~(CS) for de-noising. Unlike filtering the image in the $k$-space, which increases the SNR at the expense of the spatial resolution, CS can de-noise the data without a loss of image resolution, although at the cost of computation time. We adapted the conjugate gradients method to solve the unconstrained optimisation problem \cite{lustig2007sparse}. The equation for convex optimisation can be written as $\L = argmin \{||F m - y||_2 ^2  + \lambda ||W m ||_1\}$, where $F$ and $W$ denote fast Fourier and wavelet transformations, respectively; $m$ is a solution with the reconstructed image in  real space; $y$ is an input, which is the k-space image data. $\lambda$ is a weighting factor for the sparsity, and its value was optimised after several verification runs.
\\\indent A contrast enhancement and a better image quality can be observed after CS treatment of the image (Fig.~\ref{fig:3}~(b)). In addition, the standard deviation of the background noise has been significantly reduced, which leads to a perceivable improvement in the SNR of the image.

\section{Discussion}

\subsection{Image resolution}
To estimate the spatial resolution (z-direction) of the image shown in Fig.~\ref{fig:3}~(a) we fitted the data slices across the edges of the letter K in the image with a Fermi function. From the first derivative of the fit function, we can then obtain the image resolution, which is represented as the full width at half maximum (see the supplementary information). The average resolution obtained with this method was 1.05~mm. The lower resolution compared to the voxel size may stem from the partial volume effect induced by the inhomogeneous field gradient. We found that the background field varies from 50~to~200~nT over the sample space in our magnetically shielded room. Demagnetising the $\upmu$-metal layers used in the surrounding walls of the magnetically shielded room will alleviate the residual field gradient. We also investigated a Larmor frequency drift, but it only accounts for less than a 0.1~mm image shift over the 77~min measurement period.

\subsection{Off-resonance effect}
In the sequence shown in Fig.~\ref{fig:2}~(b), the frequency encoding gradient $G_z$ turns on immediately after the O-DNP process, and is not turned off, even while a refocusing $B_1$ pulse is applied. This simplifies the entire sequence; however, if the $B_1$ strength is weaker than the field generated by the gradient coil, a strong off-resonance effect occurs. This causes a reduction of the spin echo signal strength. We irradiated a strong $B_1$ pulse, which is estimated to be over 20~$\upmu$T from the nutation curve. Because it is circularly polarised, the $B_1$ field does not induce the Bloch-Siegert effect \cite{shim2014strong}. However, the off-resonance effect is not negligible, as the gradient field increases along the z-axis. Thus, the flip-angle of the $B_1$ pulse eventually deviates from 180$^{\circ}$, reducing the spin echo intensity. The calculation of the transverse component of the magnetisation after the refocusing $B_1$ pulse in the presence of the off-resonance field revealed that the spin echo signal gradually decreases along the z-axis and reduces to 90~$\%$ at the edges of the phantom (see the supplementary information).

\subsection{Imaging time}
The total elapsed time of 77~min for the image in Fig.~\ref{fig:3}~(a) seems not promising. However, 90~$\%$ of the measurement time is consumed by the polarisation time and the delay between measurements. There remain several methods for achieving a faster image acquisition. A rapid acquisition with relaxation enhancement (RARE) \cite{hennig1986rare} sequence can be used to accelerate the experiment to a large extent, by taking advantage of filling multiple k-space lines at the T$_2$ time scale within a single shot, thereby shortening the acquisition time. Here, T$_\mathrm{2}$ is nearly comparable with T$_\mathrm{1}$ in liquid phases, and the polarisation time used in this study is close to the value of T$_\mathrm{1}$ in the TEMPOL solution. Therefore, a RARE sequence is applicable. A time saving of at last 50~$\%$ should be possible, which may be further increased if we reduce our phase encoding time and increase the gradient fields for compensation. In addition, the delay between measurements could be further reduced if the magnetically shielded room has been optimised for eddy current suppression, such as reducing the size of the conductive plates used for RF shielding \cite{zevenhoven2014}.

\subsection{RF exposure to He dewar}
One side effect of applying O-DNP uncovered during the present study is the increase in the boil-off rate of the liquid helium stored in the dewar. Because the bottom of the dewar is located inside the birdcage coil, the superinsulation layers on the inside of the dewar are exposed to RF radiation, which causes RF absorption as the sequence runs. Placing additional RF shielding layers at the outmost surface of the dewar is expected to mitigate this issue.

\section{Conclusion}
\label{Conclusion}
It has been widely accepted that O-DNP with nitroxide radicals requires a magnetic field higher than hundreds of microtesla  for optimal efficiency\cite{Guiberteau1996}. However, we demonstrated that circularly polarised RF results in a substantially higher enhancement factor of over 150,000 at 1.2~$\upmu$T and can be implemented with a birdcage coil. The magnetic field applied for O-DNP was only 1.2~$\upmu$T ($B_p$), which is approximately 30~times lower than  34.5 $\upmu$T in our previous work\cite{lee2015magnetic}. However, the quality of the image (Fig.~\ref{fig:3}~(a)) exhibits an appreciable improvement. We expect this method to broaden the scope of Overhauser-enhanced MRI in ultra-low fields or even near zero fields. Because the magnetic field used in this work for O-DNP is no higher than a few microtesla, optically-pumped atomic magnetometers\cite{kominis2003} can be adopted in direct detection of the NMR signal as in Ref.~\cite{lee2019insitu},  with a more suppressed transient $^{87}$Rb oscillation and yet a higher nuclear polarisation. This leads to a potentially non-cryogenic ultra-low field MRI system capable of a reasonable resolution.

\section*{Acknowledgement}
\label{Acknowledgement}

This work was supported by the Development of Core Technology for Advanced Scientific Instrument (KRISS–2019–GP2019-0018) and the Development of Platform Technology for Innovative Medical Measurements Program (KRISS-2019-GP2019-0013) from the Korea Research Institute of Standards and Science. We would like to thank Yeunchul Ryu's group at the Neuroimaging Research Center of Gachon University, for the initial help and advice on building the birdcage coil.

\appendix

\section{Gyromagnetic ratio of the spin $\bm{F}$}
\label{app:a}

The calculation of of the gyromagnetic ratio of the spin $\bm{F}$ can be done similarly to that for Lande-g factor. We assume that $\gamma_F (F)$ satisfies the equation below,

\begin{equation}
\gamma_e \bm{S} + \gamma_n \bm{I} = \gamma_F (F) \bm{F}.\label{EqA1}
\end{equation}

\noindent On both sides of the Eq.~(\ref{EqA1}), we do the inner product of the spin $\bm{F}$, and replace $\bm{S}\cdot \bm{I}$ by $\frac{1}{2} [ \bm{F}^2 - (\bm{S}^2 + \bm{I}^2)]$. Then, the equation below can be obtained:

\begin{equation}
\frac{\gamma_e}{2} [ \bm{F}^2 + (\bm{S}^2 + \bm{I}^2)] + \frac{\gamma_n}{2} [ \bm{F}^2 - (\bm{S}^2 + \bm{I}^2)] = \gamma_F (F) \bm{F}^2. \label{EqA2}
\end{equation}

\noindent $\ket{F,m_F}$ are linear superposition states of the product of individual spin states as $\ket{F,m_F} = \sum_{m_S, m_I} c_{m_S, m_I} \ket{S, m_S} \ket{I, m_I}$. Thus, the equations $\langle \bm{F}^2 \rangle = F(F+1)$, $\langle \bm{S}^2 \rangle = S(S+1)$, and $\langle \bm{I}^2 \rangle = I(I+1)$ can be applied to the expectation values of both sides of the Eq.~(\ref{EqA2}).  Then, we can obtain Eq.~(\ref{Eq_gamma_F}).

\section{Calculation of the enhancement factor}
\label{app:b}

According to the Eq.~(\ref{Eq_Enhancement}), the enhancement factor is proportional to the deviations of the radical’s magnetic moment from its thermal equilibrium under a resonant oscillating field. When a transition $T_{ij}$ is induced, the deviation of the magnetic moment of the spin $\bm{F}_z$ is defined as,

\begin{equation}
\Delta \langle \gamma_F (F) \bm{F}_z \rangle_{ij} = \gamma_F (F) m_F^i \Delta n_i + \gamma_F (F) m_F^j \Delta n_j,\label{EqB1}
\end{equation}

\noindent in which $\Delta n_k = n_k - n_k^{eq}$ is the population change of the state $\ket{k}$, shown in Fig.~\ref{fig:1} (a). Populations at thermal equilibrium are determined by Boltzmann's law as $n_k^{eq} = \exp(-\frac{E_k}{k_B T})/Z$. ($Z = \sum_k \exp(-\frac{E_k}{k_B T})$). In ultralow fields, ZFS is significantly larger than Zeeman energies. Therefore, it is reasonable to assume that, at thermal equilibrium, the states in the quartet have the same population as $n_1^{eq} = n_2^{eq} = n_3^{eq} = n_4^{eq} = n(\frac{3}{2})$, and so do the states in the doublet as $n_5^{eq} = n_6^{eq} = n(\frac{1}{2})$. $n(\frac{3}{2})$ and $n(\frac{1}{2})$ satisfy the relation,

\begin{equation}
n(\frac{3}{2}) / n(\frac{1}{2}) = \exp(-\frac{3A}{2 k_B T}).
\end{equation}

\noindent With the condition that $\sum_k n_k = 1$ and high temperature approximation, the population difference between the doublet and quartet is,

\begin{equation}
\delta n = n(\frac{1}{2}) - n(\frac{3}{2}) = \frac{A}{4 k_B T}.\label{EqB3}
\end{equation}

If a transition $T_{ij}$ ($i \in \{1,2,3,4 \}, j \in \{5, 6\}$) is fully saturated, the population of $n_i$ and $n_j$ will be equal to their average $\frac{1}{2} \left[ n(\frac{3}{2}) + n(\frac{1}{2})\right]$.  Thus, $\Delta n_i = \frac{1}{2}\left[ -n(\frac{3}{2}) + n(\frac{1}{2})\right] = \frac{\delta n}{2}$ and $\Delta n_j = \frac{1}{2}\left[ n(\frac{3}{2}) - n(\frac{1}{2})\right] = -\frac{\delta n}{2}$. By putting these into Eq.~(\ref{EqB1}), we can derive the approximate equation below,

\begin{equation}
\Delta \langle \gamma_F (F) \bm{F}_z \rangle_{ij} = \frac{1}{6} \gamma_e \delta n (m_F^i + m_F^j ).\label{EqB4}
\end{equation}

In order to estimate the maximal theoretical enhancement, we assume $s=1$ and $f = 1$ in Eq.~(\ref{Eq_Enhancement}). When the dipolar interaction is dominant in Overhauser DNP, the coupling factor $\rho$ becomes positive ($\rho = 0.5$). With the thermal nuclear polarisation $I_0 = \frac{\gamma_I B}{2 k_B T}$ and Eq.~(\ref{EqB4}), the maximal enhancement can be derived as

\begin{equation}
\varepsilon_{ij}^{max} = -\frac{A}{4 \gamma_I B} \frac{m_F^i + m_F^j}{2}.\label{Eq_B5}
\end{equation}

\noindent Therefore,  the saturation of $T_{16}$ ($m_F^1 + m_F^6 = 2$) induces negative enhancement, while $T_{45}$ ($m_F^4 + m_F^5 = -2$) positive as illustrated in Fig.~\ref{fig:1} (b).

\section{Experimental methods}
\label{app:c}

\subsection{Ultra-low field MRI system}
\label{app:c-imgsys}

A schematic of the experimental setup used for our ultra-low field imaging system is shown in Fig.~\ref{fig:appc}. The coils, sensor system, and sample are located inside a magnetically shielded room. Our low noise Helium dewar contains a direct current SQUID (Supracon AG) to which a second-order gradiometer with a baselength of 50~mm and a diameter of 65~mm is connected. It is made of a Niobium wire with a diameter of 125~$\upmu$m and oriented along the y-direction. The SQUID is shielded with two superconducting layers. The sensor itself has a noise level of approximately 3~fT$\cdot$Hz$^{-\frac{1}{2}}$. With all magnetic fields, required for the measurement activated, the noise level increases to 5~fT$\cdot$Hz$^{-\frac{1}{2}}$. The main contributors are the frequency encoding gradient field $G_z$ and the imaging field $B_{m}$.
\\\indent The imaging coils include gradient coils in all three main directions $G_x$, $G_y$, and $G_z$, as well as the imaging field $B_{m}$ along the z-direction. All coils are wound on a wooden frame, which was carefully manufactured to avoid magnetic impurities. We also tried to avoid the use of magnetic materials wherever possible. The coil system consist of two frames, each separated by approximately 0.66~m, with an edge length of 1.375~m. The $G_x$ gradient coil is used for the phase encoding, whereas $G_z$ is used for the frequency encoding. Helmholtz coils along the x- and y-directions are utilised to generate a circular polarised $B_1$ pulse \cite{shim2014strong}. An additional coil along the y-direction provides a weak polarising field $B_{pol}$ of 1.2~$\upmu$T. These coils are wound on a wooden cube structure, which supports the sensor system. The imaging field $B_{m}$ is 20.69~$\upmu$T, corresponding to a Larmor frequency of 880~Hz.

\renewcommand{\thefigure}{C.\arabic{figure}}
\setcounter{figure}{0}

\begin{figure}[t!]
\centering
\includegraphics[origin=c,clip=true,width=\columnwidth]{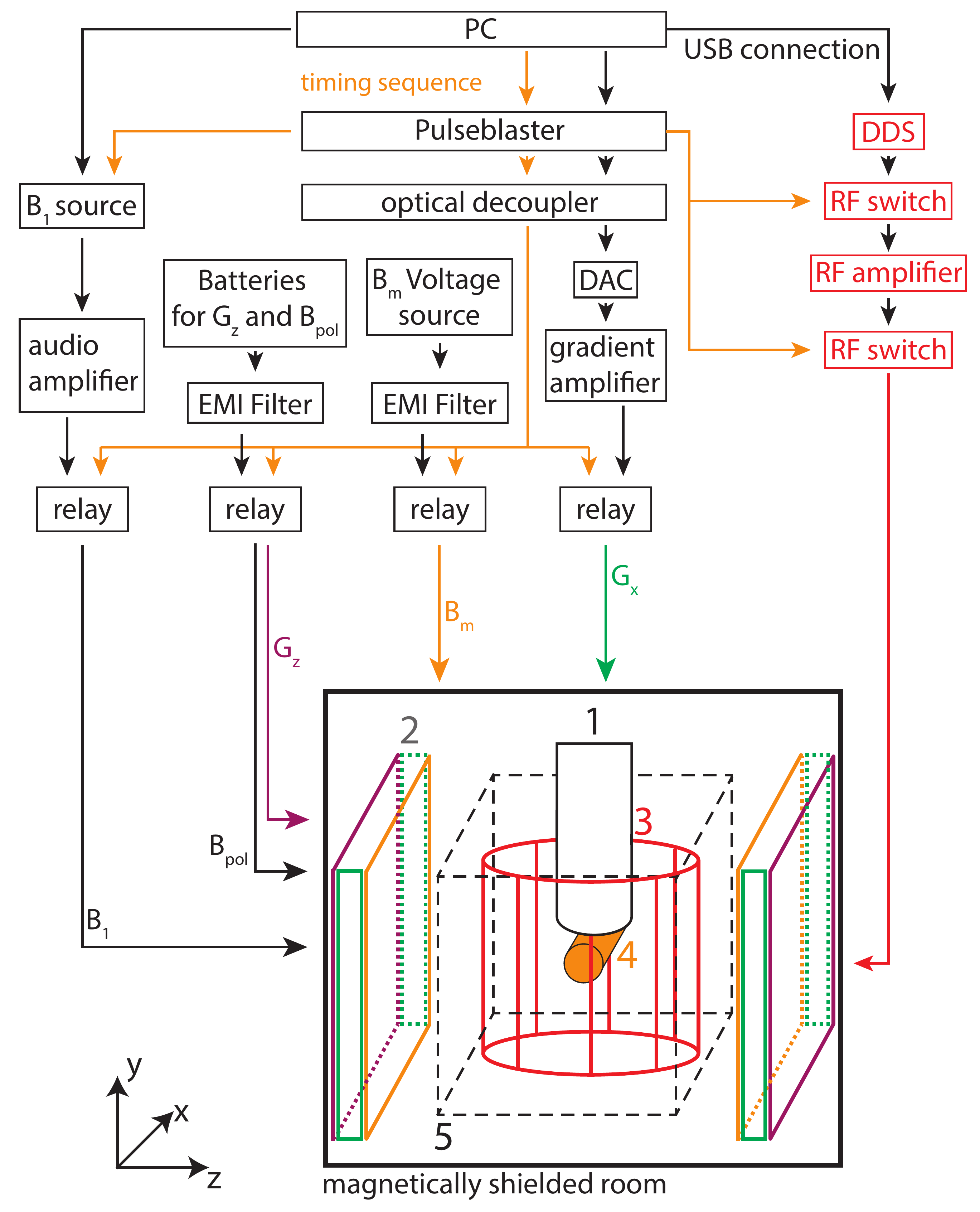}
\caption{A schematic of the ultra-low field MRI system is shown, including SQUID sensor system
		(1), gradient (PEG $G_x$, and FEG $G_z$) and imaging field $B_m$ coils
		(2), birdcage coil
		(3), sample
		(4), the tipping coil $B_1$ in x- and y-direction and $B_p$ coil for a polarisation field along the y-direction (5).}
\label{fig:appc}
\end{figure}

The necessary digital triggers are generated using a timing signal generator (PulseBlaster, Spincore), and fed to an optical decoupler array. The relay switches are then addressed using optical fibres. They are required to decouple the coils from the current sources, while the signal acquisition runs, reducing the noise level. We used batteries for the relatively weak polarising field $B_{pol}$ and the frequency encoding gradient $G_z$ to reduce the power-line noise. The imaging field $B_m$ is generated using a low noise current source (Keithley PWS4323) for stability.
\\\indent To generate a circularly polarised $B_1$ pulse, we used a two-channel frequency generator (Agilent 33500B) programmed for arbitrary waveforms for accurate pulse shapes. Both channels apply a signal with a frequency corresponding to a Larmor frequency of 880~Hz. The outputs were then amplified using a two-channel audio amplifier and fed through relays, which decouples the signals during the acquisition time. We attached a very detailed schematic of the experimental setup in the supplementary information for the interested readers.

\subsection{Imaging sequence}
\label{app:c-seq}
A spin-echo sequence was used for imaging, as shown in Fig.~\ref{fig:2}~(c). Unlike a conventional sequence, the initial tipping pulse is omitted because $B_p$, along which the enhanced nuclear magnetisation aligns, is orthogonal to $B_m$. Thus, only a refocusing pulse is necessary for a spin echo. We use a polarisation time $t_{pol}$ of 2000~ms, a phase encoding time $t_ph$ of 100~ms, and a read-out time $t_{acq}$ of 200~ms. The echo time is approximately 200~ms, which is much shorter than the T$_2$ time of the sample. The resulting signal loss is approximately 10~$\%$. The duration for the $B_1$ pulse is 0.58~ms. To suppress the effects of eddy currents induced in the walls of the magnetically shielded room, we applied an additional delay of 1000~ms ($t_d$) between each measurement, which is relatively long because the shielding walls of the magnetically shielded room are not optimised for suppressing the eddy current effects and last a fairly long time.

\subsection{Circularly polarised RF}
\label{app:c-circ}
Because hyperfine coupling ($A$) between the electron and $^{14}$N nuclear spins is approximately 48~MHz, zero-field splitting ($= 3A/2$) occurs at near 72~MHz \cite{lee2015dynamic}, which is the frequency of the circularly polarised RF field used. Two phase-synced sinusoidal waves are generated using an in-house built direct digital synthesizer (DDS, AD9958). The phase shift between them is adjusted for circular polarisation. The in-phase and quadrature RF signals are amplified separately up to 18~W. We monitor the amplified RF signals using two directional couplers (Pulsar C30-102-481/2N) to confirm the power and phase consistency. High-power RF switches positioned after the amplifier reduce the disturbances of the SQUID sensor during data acquisition. The amplified RF signals, in-phase and quadrature, are fed to the birdcage coil inputs.

\subsection{Birdcage coil}
\label{app:c-bc}
A single birdcage coil can generate circularly polarised RF fields by driving the coil with in-phase and quadrature RF signals aligned perpendicularly. It is more suitable than a quadrature surface coil owing to its high spatial homogeneity in the coil volume and ease of access. Among the three types of birdcage coils, we have chosen the high-pass type. The coil consists of legs, also referred to as rungs, which are equally spaced around the surface of the cylinder. They connect to one ring on either end of the cylinder. For the high-pass type, the capacitors are located on the ring, between the segments of the end points of the rungs. We used 5~mm wide, thin circuit board strips to build the coil, allowing it to be sufficiently flexible to follow the curvature of the cylinder body. The input circuit is balanced, using two matching capacitors C$_\mathrm{m}$.
\\ \indent To suppress common mode currents through the coax cables connected to the input circuits, we added cable traps tuned to 72~MHz at every half-wavelength position. A magnetic field probe\cite{ediss2003probing} was built for the tuning of the cable traps. Additionally, ferrite cores were positioned at the end of the coax cables connected to the RF amplifier. The developed birdcage coil electrical schematic is shown in Fig.~\ref{fig:2}~(c). The coil has an outer diameter of 15~cm with eight legs, each 15~cm long. We chose a ratio close to 1:1 for diameter-to-rung lengths to maximise the homogeneity of our coil \cite{zanche2007birdcage}. The $"$Birdcage Builder$"$ software was used to calculate the required capacitance \cite{chin1998birdcage}. Adjustments of all capacitors for tuning and matching were applied with the dewar reaching inside the birdcage coil, because metallic components in the dewar strongly disturb the resonance modes. Fine adjustments prior to each measurement can be conducted using the variable capacitors (Voltronics V9000). If one disregards the end ring modes, the main mode has the highest frequency \cite{leifer1997resonant, doty1999practical,vullo1992experimental}.
\\ \indent The crosstalk was adjusted using $C_{45}$, which defines the independence of the two input channels. If the two channels have a higher crosstalk, the resulting field will converge to a linear polarised field rather than the intended circular polarised field. We achieved a crosstalk of 13~$\%$ in amplitude, which was measured by placing a pick-up coil at the centre of the birdcage coil and rotating it.


\section*{References}
\bibliographystyle{elsarticle-num}
\bibliography{bibliography2}

\begin{thebibliography}{10}
\expandafter\ifx\csname url\endcsname\relax
  \def\url#1{\texttt{#1}}\fi
\expandafter\ifx\csname urlprefix\endcsname\relax\def\urlprefix{URL }\fi
\expandafter\ifx\csname href\endcsname\relax
  \def\href#1#2{#2} \def\path#1{#1}\fi

\bibitem{volegov2004simultaneous}
P.~Volegov, A.~N. Matlachov, M.~A. Espy, J.~S. George, R.~H. Kraus,
  Simultaneous magnetoencephalography and squid detected nuclear mr in
  microtesla magnetic fields, Magn. Reson. Med. 52~(3) (2004) 467--470.

\bibitem{vesanen2013hybrid}
P.~T. Vesanen, J.~O. Nieminen, K.~C. Zevenhoven, J.~Dabek, L.~T. Parkkonen,
  A.~V. Zhdanov, J.~Luomahaara, J.~Hassel, J.~Penttil{\"a}, J.~Simola, et~al.,
  Hybrid ultra-low-field mri and magnetoencephalography system based on a
  commercial whole-head neuromagnetometer, Magn. Reson. Med. 69~(6) (2013)
  1795--1804.

\bibitem{McDermott2002}
R.~McDermott, A.~H. Trabesinger, M.~Mück, E.~L. Hahn, A.~Pines, J.~Clarke,
  Liquid-state nmr and scalar couplings in microtesla magnetic fields, Science
  295~(5563) (2002) 2247.

\bibitem{Ledbetter2009}
M.~P. Ledbetter, C.~W. Crawford, A.~Pines, D.~E. Wemmer, S.~Knappe,
  J.~Kitching, D.~Budker, Optical detection of nmr j-spectra at zero magnetic
  field, J. Magn. Reson. 199~(1) (2009) 25--29.

\bibitem{Theis2013}
T.~Theis, J.~W. Blanchard, M.~C. Butler, M.~P. Ledbetter, D.~Budker, A.~Pines,
  Chemical analysis using j-coupling multiplets in zero-field nmr, Chem. Phys.
  Lett. 580 (2013) 160--165.

\bibitem{Sarah2012}
B.~Sarah, H.~Michael, M.~Michael, M.~Whittier, W.~Travis, M.~Michael, C.~Kevin,
  K.~Kyle, S.~Jeffry, C.~John, Measurements of t1 relaxation in ex vivo
  prostate tissue at 132 $\mu \mathrm{T}$, Magn. Reson. Med. 67~(4) (2012)
  1138--1145.

\bibitem{Moessle2006}
M.~Mossle, S.-I. Han, W.~R. Myers, S.-K. Lee, N.~Kelso, M.~Hatridge, A.~Pines,
  J.~Clarke, Squid-detected microtesla mri in the presence of metal, J. Magn.
  Reson. 179~(1) (2006) 146--151.

\bibitem{Kim2014}
K.~Kim, S.-J. Lee, C.~S. Kang, S.-m. Hwang, Y.-H. Lee, K.-K. Yu, Toward a brain
  functional connectivity mapping modality by simultaneous imaging of coherent
  brainwaves, NeuroImage 91 (2014) 63--69.

\bibitem{korber2016squids}
R.~K{\"o}rber, J.-H. Storm, H.~Seton, J.~P. M{\"a}kel{\"a}, R.~Paetau,
  L.~Parkkonen, C.~Pfeiffer, B.~Riaz, J.~F. Schneiderman, H.~Dong, et~al.,
  Squids in biomagnetism: a roadmap towards improved healthcare, Supercond.
  Sci. Technol. 29~(11) (2016) 113001.

\bibitem{Griffith2009}
W.~C. Griffith, M.~D. Swallows, T.~H. Loftus, M.~V. Romalis, B.~R. Heckel,
  E.~N. Fortson, Improved limit on the permanent electric dipole moment of
  $^{199}\mathrm{Hg}$, Phys. Rev. Lett. 102~(10) (2009) 101601.

\bibitem{greenberg1998application}
Y.~S. Greenberg, Application of superconducting quantum interference devices to
  nuclear magnetic resonance, Reviews of Modern Physics 70~(1) (1998) 175.

\bibitem{McDermott2004}
R.~McDermott, S.~Lee, B.~t. Haken, A.~H. Trabesinger, A.~Pines, J.~Clarke,
  Microtesla mri with a superconducting quantum interference device, Proc.
  Natl. Acad. Sci. U.S.A. 101~(21) (2004) 7857.

\bibitem{kominis2003}
I.~K. Kominis, T.~W. Kornack, J.~C. Allred, M.~V. Romalis, A subfemtotesla
  multichannel atomic magnetometer, Nature 422 (2003) 596.

\bibitem{Budker2007}
D.~Budker, M.~Romalis, Optical magnetometry, Nat. Phys. 3 (2007) 227.

\bibitem{Hwang2011}
S.-M. Hwang, K.~Kim, C.~Seok~Kang, S.-J. Lee, Y.-H. Lee, Effective cancellation
  of residual magnetic interference induced from a shielded environment for
  precision magnetic measurements, Appl. Phys. Lett. 99~(13) (2011) 132506.

\bibitem{nieminen2011avoiding}
J.~O. Nieminen, P.~T. Vesanen, K.~C. Zevenhoven, J.~Dabek, J.~Hassel,
  J.~Luomahaara, J.~S. Penttil{\"a}, R.~J. Ilmoniemi, Avoiding eddy-current
  problems in ultra-low-field mri with self-shielded polarizing coils, J. Magn.
  Reson. 212~(1) (2011) 154--160.

\bibitem{zevenhoven2014}
K.~C.~J. Zevenhoven, S.~Busch, M.~Hatridge, F.~Öisjöen, R.~J. Ilmoniemi,
  J.~Clarke, Conductive shield for ultra-low-field magnetic resonance imaging:
  Theory and measurements of eddy currents, Journal of Applied Physics 115~(10)
  (2014) 103902.
\newblock \href {http://dx.doi.org/10.1063/1.4867220}
  {\path{doi:10.1063/1.4867220}}.

\bibitem{storm2016modular}
J.-H. Storm, D.~Drung, M.~Burghoff, R.~K{\"o}rber, A modular, extendible and
  field-tolerant multichannel vector magnetometer based on current sensor
  squids, Supercond. Sci. Technol. 29~(9) (2016) 094001.

\bibitem{lee2015dynamic}
S.-J. Lee, J.~H. Shim, K.~Kim, K.~K. Yu, S.-m. Hwang, Dynamic nuclear
  polarization in the hyperfine-field-dominant region, J. Magn. Reson. 255
  (2015) 114--121.

\bibitem{Zotev2010}
V.~S. Zotev, T.~Owens, A.~N. Matlashov, I.~M. Savukov, J.~J. Gomez, M.~A. Espy,
  Microtesla mri with dynamic nuclear polarization, J. Magn. Reson. 207~(1)
  (2010) 78--88.

\bibitem{lurie1988proton}
D.~J. Lurie, D.~M. Bussell, L.~H. Bell, J.~R. Mallard, Proton-electron double
  magnetic resonance imaging of free radical solutions, Journal of magnetic
  resonance 76 (1988) 366--370.

\bibitem{Hoevener2013}
J.-B. Hoevener, N.~Schwaderlapp, T.~Lickert, S.~B. Duckett, R.~E. Mewis,
  L.~A.~R. Highton, S.~M. Kenny, G.~G.~R. Green, D.~Leibfritz, J.~G. Korvink,
  J.~Hennig, D.~von Elverfeldt, A hyperpolarized equilibrium for magnetic
  resonance, Nat. Commun. 4 (2013) 2946.

\bibitem{Buckenmaier2017}
K.~Buckenmaier, M.~Rudolph, C.~Back, T.~Misztal, U.~Bommerich, P.~Fehling,
  D.~Koelle, R.~Kleiner, H.~A. Mayer, K.~Scheffler, J.~Bernarding, M.~Plaumann,
  Squid-based detection of ultra-low-field multinuclear nmr of substances
  hyperpolarized using signal amplification by reversible exchange, Sci. Rep.
  7~(1) (2017) 13431.

\bibitem{slichter2013principles}
C.~P. Slichter, Principles of magnetic resonance, Vol.~1, Springer Science \&
  Business Media, 2013.

\bibitem{krishna2002overhauser}
M.~C. Krishna, S.~English, K.~Yamada, J.~Yoo, R.~Murugesan, N.~Devasahayam,
  J.~A. Cook, K.~Golman, J.~H. Ardenkjaer-Larsen, S.~Subramanian, et~al.,
  Overhauser enhanced magnetic resonance imaging for tumor oximetry:
  coregistration of tumor anatomy and tissue oxygen concentration, Proceedings
  of the National Academy of Sciences 99~(4) (2002) 2216--2221.

\bibitem{sarracanie2014high}
M.~Sarracanie, B.~D. Armstrong, J.~Stockmann, M.~S. Rosen, High speed 3d
  overhauser-enhanced mri using combined b-ssfp and compressed sensing,
  Magnetic resonance in medicine 71~(2) (2014) 735--745.

\bibitem{waddington2018overhauser}
D.~E. Waddington, M.~Sarracanie, N.~Salameh, F.~Herisson, C.~Ayata, M.~S.
  Rosen, An overhauser-enhanced-mri platform for dynamic free radical imaging
  in vivo, NMR in Biomedicine 31~(5) (2018) e3896.

\bibitem{Guiberteau1996}
T.~Guiberteau, D.~Grucker, Epr spectroscopy by dynamic nuclear polarization in
  low magnetic field, Journal of Magnetic Resonance, Series B 110~(1) (1996)
  47--54.

\bibitem{lee2015magnetic}
S.-J. Lee, J.~H. Shim, K.~Kim, K.~K. Yu, S.-m. Hwang, Magnetic resonance
  imaging without field cycling at less than earth's magnetic field, Appl.
  Phys. Lett. 106~(10) (2015) 103702.

\bibitem{liebgott2003proton}
T.~Liebgott, H.~Li, Y.~Deng, J.~L. Zweier, Proton electron double resonance
  imaging (pedri) of the isolated beating rat heart, Magnetic Resonance in
  Medicine: An Official Journal of the International Society for Magnetic
  Resonance in Medicine 50~(2) (2003) 391--399.

\bibitem{nicholson1994application}
I.~Nicholson, D.~Lurie, F.~Robb, The application of proton-electron
  double-resonance imaging techniques to proton mobility studies, Journal of
  Magnetic Resonance, Series B 104~(3) (1994) 250--255.

\bibitem{nieminen2010solving}
J.~O. Nieminen, R.~J. Ilmoniemi, Solving the problem of concomitant gradients
  in ultra-low-field mri, J. Magn. Reson. 207~(2) (2010) 213--219.

\bibitem{lustig2007sparse}
M.~Lustig, D.~Donoho, J.~M. Pauly, Sparse mri: The application of compressed
  sensing for rapid mr imaging, Magn. Reson. Med. 58~(6) (2007) 1182--1195.

\bibitem{shim2014strong}
J.~H. Shim, S.-J. Lee, K.-K. Yu, S.-M. Hwang, K.~Kim, Strong pulsed excitations
  using circularly polarized fields for ultra-low field nmr, J. Magn. Reson.
  239 (2014) 87--90.

\bibitem{hennig1986rare}
J.~Hennig, A.~Nauerth, H.~Friedburg, Rare imaging: a fast imaging method for
  clinical mr, Magnetic resonance in medicine 3~(6) (1986) 823--833.

\bibitem{lee2019insitu}
H.~J. Lee, S.-J. Lee, J.~H. Shim, H.~S. Moon, K.~Kim, In-situ
  overhauser-enhanced nuclear magnetic resonance at less than 1 $\mu
  \mathrm{T}$ using an atomic magnetometer, J. Magn. Reson. 300 (2019)
  149--152.

\bibitem{ediss2003probing}
R.~Ediss, P.~Semiconductors, Probing the magnetic field probe, EMC \&
  Compliance Journal~(47).

\bibitem{zanche2007birdcage}
N.~D. Zanche, Birdcage volume coil design, eMagRes.

\bibitem{chin1998birdcage}
C.~Chin, C.~M. Collins, S.~Li, B.~J. Dardzinski, M.~B. Smith, Birdcagebuilder:
  Design of specified geometry birdcage coils with desired current pattern and
  resonant frequency, Concepts in Magnetic Resonance 15~(2) (1998) 156--163.

\bibitem{leifer1997resonant}
M.~C. Leifer, Resonant modes of the birdcage coil, J. Magn. Reson. 124~(1)
  (1997) 51--60.

\bibitem{doty1999practical}
F.~D. Doty, G.~Entzminger~Jr, C.~D. Hauck, J.~P. Staab, Practical aspects of
  birdcage coils, J. Magn. Reson. 138~(1) (1999) 144--154.

\bibitem{vullo1992experimental}
T.~Vullo, R.~T. Zipagan, R.~Pascone, J.~P. Whalen, P.~T. Cahill, Experimental
  design and fabrication of birdcage resonators for magnetic resonance imaging,
  Magn. Reson. Med. 24~(2) (1992) 243--252.

\end{thebibliography}

\end{document}